\documentclass[traditabstract]{aa}

\usepackage{txfonts}

\usepackage{natbib}
\usepackage{graphicx}

\begin{document}
\title{WASP-26b: A 1-Jupiter-mass planet around an early-G-type star}
\author
{B.~Smalley\inst{1},
D.~R.~Anderson\inst{1},
A.~Collier Cameron\inst{2},
M.~Gillon\inst{3,4},
C.~Hellier\inst{1},
T.~A.~Lister\inst{5},
P.~F.~L.~Maxted\inst{1},
D.~Queloz\inst{4},
A.~H.~M.~J.~Triaud\inst{4},
R.~G.~West\inst{6},
S.~J.~Bentley\inst{1},
B.~Enoch\inst{2},
F.~Pepe\inst{4},
D.~L.~Pollacco\inst{7},
D.~Segransan\inst{4},
A.~M.~S.~Smith\inst{1},
J.~Southworth\inst{1},
S.~Udry\inst{4},
P.~J.~Wheatley\inst{8},
P.~L.~Wood\inst{1},
J.~Bento\inst{8}
}

\authorrunning{B. Smalley et al.}
\titlerunning{WASP-26b}

\offprints{Barry Smalley \email{bs@astro.keele.ac.uk}}

\institute{
        Astrophysics Group, Keele University,
        Staffordshire, ST5 5BG, United Kingdom
\and
School of Physics and Astronomy, University of St. Andrews,
North Haugh, Fife, KY16 9SS, UK
\and
Institut d'Astrophysique et de G\'{e}ophysique, Universit\'{e}
de Li\`{e}ge, All\'{e}e du 6 A\^{u}t, 17, Bat. B5C, Li\`{e}ge 1, Belgium
\and
Observatoire de Gen\`{e}ve, Universit\'{e} de Gen\`{e}ve, 51
Chemin des Maillettes, 1290 Sauverny, Switzerland
\and
Las Cumbres Observatory, 6740 Cortona Dr. Suite 102,
Santa Barbara, CA 93117, USA
\and
Department of Physics and Astronomy, University of
Leicester, Leicester, LE1 7RH, UK
\and
Astrophysics Research Centre, School of Mathematics
\& Physics, Queen's University, University Road, Belfast,
BT7 1NN, UK
\and
Department of Physics, University of Warwick, Coventry CV4 7AL, UK
}

\date{Received date / accepted date}

\abstract{We report the discovery of WASP-26b, a moderately over-sized
Jupiter-mass exoplanet transiting its 11.3-magnitude early-G-type host star
(1SWASP\,J001824.70-151602.3; TYC 5839-876-1) every 2.7566~days. A simultaneous
fit to transit photometry and radial-velocity measurements yields a planetary
mass of $1.02 \pm 0.03$~$M_{\rm Jup}$ and radius of $1.32 \pm 0.08$~$R_{\rm
Jup}$. The host star, WASP-26, has a mass of $1.12 \pm 0.03$~$M_{\sun}$ and a
radius of $1.34 \pm 0.06$~$R_{\sun}$ and is in a visual double with a fainter
K-type star. The two stars are at least a common-proper motion pair with a
common distance of around $250 \pm 15$~pc and an age of $6 \pm 2$~Gy.}


\maketitle

\section{Introduction}

Most of the known exoplanets have been discovered using the radial velocity
technique \citep{1995Natur.378..355M}. However, in recent years an increasing
number have been discovered using the transit technique, via ground-based and
space-based survey projects. Transiting exoplanets allow parameters such as the
mass, radius, and density to be accurately determined, as well as their
atmospheric properties to be studied during their transits and occultations
\citep{2005ApJ...626..523C,2009MNRAS.394..272S,2009IAUS..253...99W}.

The SuperWASP project has a robotic observatory on La Palma in the Canary
Islands and another in Sutherland in South Africa. The wide angle survey is
designed to find planets around relatively bright stars in the $V$-magnitude
range $9\sim13$. A detailed description is given in \citet{2006PASP..118.1407P}.

In this paper we report the discovery of WASP-26b, a Jupiter-mass planet in
orbit around its $V = 11.3$~{mag.} host star 1SWASP\,J001824.70-151602.3 in the
constellation Cetus. We present the SuperWASP-South discovery photometry,
together with follow-up optical photometry and radial velocity measurements.

\section{Observations}

\subsection{SuperWASP photometry}
\label{WASP_Phot}

\begin{figure} 

\includegraphics[height=\columnwidth,,angle=-90]{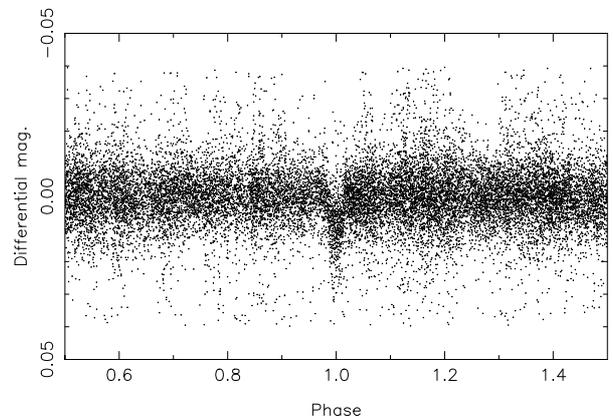}

\caption{SuperWASP photometry of WASP-26 folded on the orbital of period of
2.7566~days.}

\label{WASP-phot} 
\end{figure}

The host star WASP-26 (1SWASP\,J001824.70-151602.3; TYC\,5839-876-1) was within
two fields observed by SuperWASP-South during the 2008 and 2009 observing
seasons, covering the intervals 2008 June 30 to November 17 and 2009 June 28 to
November 17. A total of 18\,807 data points were obtained. The
pipeline-processed data were de-trended and searched for transits using the
methods described in \citet{2006MNRAS.373..799C}, yielding a detection of a
periodic, transit-like signature with a period of 2.7566~days and a depth of
0.009~magnitudes (Fig.~\ref{WASP-phot}).

There is a second star (1SWASP\,J001825.25-151613.8; USNO-B1\,0747-0003869),
$\sim$2.5 magnitudes fainter, 15{\arcsec} from WASP-26. Both stars are contained
within the 3.5-pixel ($\equiv48\arcsec$) reduction aperture. Hence, from the
SuperWASP photometry alone, we could not be totally sure that WASP-26 was the
star varying and not the fainter one in deep eclipse. Targeted photometry was
obtained to confirm that the transit signature was indeed from WASP-26 (see
Sect.~\ref{FT_Phot}).

\subsection{Spectroscopic observations with CORALIE}
\label{CORALIE}

\begin{figure} 
\includegraphics[height=\columnwidth,angle=-90]{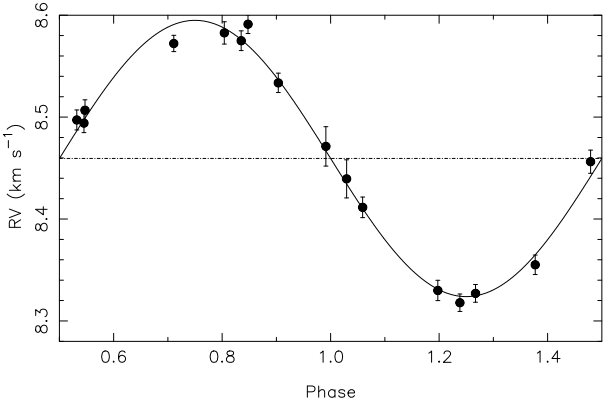}
\caption{Radial Velocity curve of WASP-26.
The solid line is the best-fitting MCMC solution.
The centre-of-mass velocity, $\gamma$, is indicated by the dashed line.} 
\label{RV} 
\end{figure}

\begin{figure}
\includegraphics[height=\columnwidth,angle=-90]{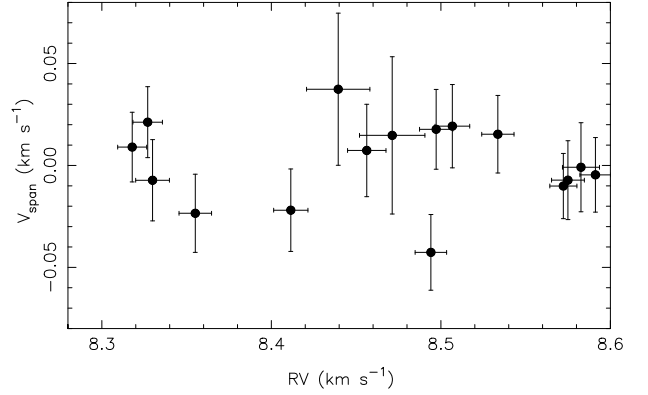}
\caption{Line bisectors ($V_{span}$) as a function of RV for WASP-26.
Bisector uncertainties of twice the RV uncertainties have been adopted.
There is no correlation between $V_{span}$ and the stellar RV.}
\label{RV-BS}
\end{figure}

Spectroscopic observations were obtained with the CORALIE spectrograph on the
Swiss 1.2m telescope. The data were processed using the standard pipeline
\citep{1996A&AS..119..373B,2000A&A...354...99Q,2002A&A...388..632P}. A total of
16 radial velocity (RV) measurements were made between 2009 June 19 and 2009
September 22 (Table~\ref{rv-data}). The resulting radial velocity curve is shown
in Fig.~\ref{RV}. The absence of any correlation with RV of the line bisector
spans ($V_{span}$) in Fig.~\ref{RV-BS} indicates that the RV variations are not
due to an unresolved eclipsing binary \citep{2001A&A...379..279Q}.

\begin{table} 
\caption{Radial velocity (RV) and line bisector spans ($V_{span}$)
measurements for WASP-26 obtained by CORALIE spectra.} 
\label{rv-data} 
\centering
\begin{tabular}{lll} \hline\hline
BJD--2\,450\,000 & RV (km\,s$^{-1}$) & $V_{span}$ (km\,s$^{-1}$) \\ \hline
5001.927439 & 8.59117 $\pm$ 0.00915 &         $-$0.00461 \\
5008.900119 & 8.35510 $\pm$ 0.00959 &         $-$0.02346 \\
5011.938063 & 8.45626 $\pm$ 0.01134 & \phantom{+}0.00736 \\
5012.833505 & 8.58264 $\pm$ 0.01092 &         $-$0.00088 \\
5013.918513 & 8.32991 $\pm$ 0.00997 &         $-$0.00727 \\
5036.934802 & 8.50676 $\pm$ 0.01022 & \phantom{+}0.01928 \\
5037.917112 & 8.53360 $\pm$ 0.00951 & \phantom{+}0.01532 \\
5038.919423 & 8.32703 $\pm$ 0.00871 & \phantom{+}0.02124 \\
5040.915661 & 8.47129 $\pm$ 0.01929 & \phantom{+}0.01476 \\
5042.899210 & 8.57224 $\pm$ 0.00800 &         $-$0.01009 \\
5068.668524 & 8.41145 $\pm$ 0.01011 &         $-$0.02195 \\
5070.807797 & 8.57495 $\pm$ 0.00967 &         $-$0.00721 \\
5072.729127 & 8.49720 $\pm$ 0.00979 & \phantom{+}0.01774 \\
5076.856687 & 8.43943 $\pm$ 0.01866 & \phantom{+}0.03740 \\
5094.818292 & 8.49411 $\pm$ 0.00929 &         $-$0.04264 \\
5096.728728 & 8.31789 $\pm$ 0.00855 & \phantom{+}0.00901 \\
\hline
\end{tabular} 
\end{table} 

\subsection{Photometry with Faulkes Telescopes}
\label{FT_Phot}

WASP-26 was observed photometrically on 2009 November 18 using the 2.0-m Faulkes
Telescope South (FTS, Siding Spring, Australia) and on 2009 December 2 using the
2.0-m Faulkes Telescope North (FTN, Maui, Hawai'i), both telescopes being
operated by LCOGT. In both cases, a new Spectral CCD
imager\footnote{http://www.specinst.com} was used along with a Pan-STARRS-$z$
filter. The Spectral instrument contains a Fairchild CCD486 back-illuminated
$4096 \times 4096$ pixel CCD which was binned $2 \times 2$ giving 0.303{\arcsec}
pixels and a field of view of $10\arcmin \times 10\arcmin$. The telescope was
defocussed a small amount during the observations to prevent saturation in the
core of the PSF but not by enough to cause blending problems with the close
($\sim 15$\arcsec) fainter companion.

\begin{figure} 
\includegraphics[height=\columnwidth,angle=-90]{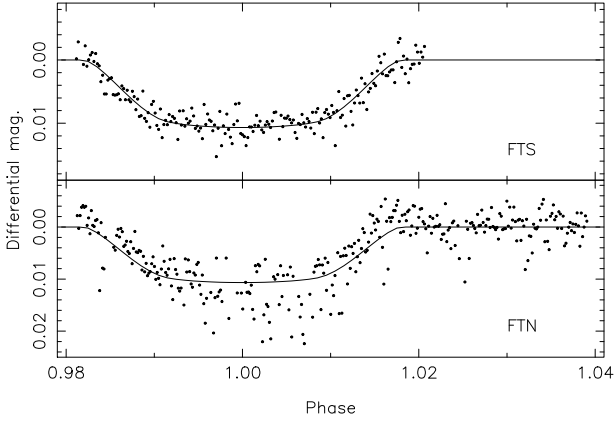}

\caption{Faulkes Telescope photometry of transits of WASP-26b. The upper plot is
from FTS on 2009 November 18 and the lower one from FTN 2009 December 2. Note
that the FTN lightcurve is somewhat affect by cloud. The solid line is the
best-fit MCMC solution.}

\label{FT-phot} 
\end{figure}

The frames were pre-processed through the WASP Pipeline
\citep{2006PASP..118.1407P} to perform overscan correction, bias subtraction and
flat-fielding. The DAOPHOT photometry package within IRAF\footnote{IRAF is
distributed by the National Optical Astronomy Observatories, which are operated
by the Association of Universities for Research in Astronomy, Inc., under
cooperative agreement with the National Science Foundation.} was used to perform
object detection and aperture photometry using aperture radii of 9 and 10 pixels
for the FTS and FTN data, respectively. Differential photometry was performed
relative to several comparison stars within the field of view. Figure~\ref{FT-phot}
shows the transit photometry. The large scatter in the FTN observations suggests
that they were somewhat affected by cloud.

\section{Spectral analysis of host star}
\label{WASP-params}

The individual CORALIE spectra of WASP-26 were co-added to produce a single
spectrum with an average S/N of around 70:1. The standard pipeline reduction
products were used in the analysis.

The analysis was performed using the methods given in
\citet{2009A&A...496..259G} and \citet{2005MSAIS...8..130S}. The H$_\alpha$ line
was used to determine the effective temperature ($T_{\rm eff}$), while the
Na\,{\sc i}~D and Mg\,{\sc i}~b lines were used as surface gravity ($\log g$)
diagnostics. The parameters obtained from the analysis are listed in
Table~\ref{WASP26-params}. The elemental abundances were determined from
equivalent width measurements of several clean and unblended lines. The quoted
error estimates include that given by the uncertainties in $T_{\rm eff}$, $\log
g$ and $\xi_{\rm t}$, as well as the scatter due to measurement and atomic data
uncertainties.

The projected stellar rotation velocity ($v \sin i$) was determined by fitting
the profiles of several unblended Fe~{\sc i} lines. A value for macroturbulence
($v_{\rm mac}$) of $4.1 \pm 0.3$~km\,s$^{-1}$ was assumed, based on the
tabulation by \citet{2008oasp.book.....G}, and an instrumental FWHM of $0.11 \pm
0.01$~{\AA}, determined from the telluric lines around 6300{\AA}. A best fitting
value of $v \sin i = 2.4 \pm 1.3$~km\,s$^{-1}$ was obtained.

\begin{table}[h]
\caption{Stellar parameters of WASP-26.}
\centering
\begin{tabular}{ll} \hline\hline
Parameter      & Value \\ \hline
RA (J2000.0)   & 00h18m24.70s \\
Dec (J2000.0)  & $-$15{\degr}16{\arcmin}02.3{\arcsec} \\
$V$ {mag.}     & 11.3 \\
$T_{\rm eff}$  & 5950 $\pm$ 100 K \\
$\log g$       & 4.3 $\pm$ 0.2 \\
$\xi_{\rm t}$  & 1.2 $\pm$ 0.1 km\,s$^{-1}$ \\
$v \sin i$     & 2.4 $\pm$ 1.3 km\,s$^{-1}$ \\
{[Fe/H]}       &$-$0.02 $\pm$ 0.09 \\
{[Si/H]}       &  +0.07 $\pm$ 0.09 \\
{[Ca/H]}       &  +0.08 $\pm$ 0.12 \\
{[Ti/H]}       &  +0.03 $\pm$ 0.06 \\
{[Ni/H]}       &   0.00 $\pm$ 0.06 \\
log A(Li)      &   1.90 $\pm$ 0.12 \\
Spectral Type  & G0 \\
$M_{\star}$    & 1.12 $\pm$ 0.03  $M_{\sun}$ \\
$R_{\star}$    & 1.34 $\pm$ 0.06  $R_{\sun}$ \\
$\rho_{\star}$ & 0.47 $\pm$ 0.06 $\rho_{\sun}$ \\
\hline
\end{tabular}
\label{WASP26-params}

\tablefoot{The spectral type was estimated from the Table~B.1 of
\citet{2008oasp.book.....G} and $M_{\star}$, $R_{\star}$ and $\rho_{\star}$ are
from the MCMC analysis (Sect.~\ref{MCMC}).}

\end{table}

The lithium abundance in WASP-26 implies an age of several ($\ga$5) Gy according
to \citet{2005A&A...442..615S}. However, recent work by
\citet{2009Natur.462..189I} suggests that lithium-depletion is enhanced in stars
with planets, making the use of lithium as an age indicator less reliable. The
measured $v \sin i$ of WASP-26 implies a rotational period of $P_{\rm rot}
\simeq 25^{+30}_{-10}$~days, which yields gyrochronological age of $\sim
7^{+24}_{-4}$~Gy using the relation of \citet{2007ApJ...669.1167B}. A search for
rotational modulation due to starspots yielded a null result, which is
consistent with the lack of stellar activity indicated by the absence of calcium
H+K emission in the CORALIE spectra.

\section{Planetary system parameters}
\label{MCMC}

To determine the planetary and orbital parameters the CORALIE radial velocity
measurements were combined with the photometry from WASP and Faulkes Telescopes
in a simultaneous fit using the Markov Chain Monte Carlo (MCMC) technique. The
details of this process are described in \citet{2007MNRAS.380.1230C} and
\citet{2008MNRAS.385.1576P}. Four sets of solutions were used: with and without
the main-sequence mass-radius constraint for both circular and floating
eccentricity orbits.

With the main-sequence constraint imposed and the eccentricity floating, a value
of $e = 0.036^{+0.031}_{-0.027}$ is found, which is significant only at the 22\%
level \citep{1971AJ.....76..544L}. The fit is indistinguishable from that with
$e = 0$, and has very little effect on the planetary radius determined. Hence, a
circular planetary orbit solution was adopted. Relaxing the main-sequence
mass-radius constraint increased the impact parameter ($b$) and the stellar and
planetary radii. Table~\ref{WASP-26b} gives the best-fit MCMC solution. The
stellar mass and radius determined by the MCMC analysis (given in
Table~\ref{WASP26-params}) are consistent with the slightly evolved nature of
WASP-26 (see Sect.~\ref{WASP26-double}).

\begin{table} 
\caption{System parameters for WASP-26b.} 
\label{WASP-26b} 
\centering
\begin{tabular}{ll} \hline\hline
Parameter & Value \\ \hline 
Orbital period, $P$  & 2.75660 $\pm$ 0.00001 days \\
Transit epoch (HJD), $T_0$ & 2455123.6379 $\pm$ 0.0005 \\
Transit duration, $T_{14}$ & 0.098 $\pm$ 0.002 days \\
$(R_{\rm P}/R_{*})^{2}$ & 0.0103 $\pm$ 0.0004 \\
Impact parameter, $b$ & 0.83 $\pm$ 0.02 \\
Reflex velocity, $K_1$ & 0.1355 $\pm$ 0.0035 km\,s$^{-1}$ \\
Centre-of-mass velocity, $\gamma$ & 8.4594 $\pm$ 0.0002 km\,s$^{-1}$ \\
Orbital separation, $a$ & 0.0400 $\pm$ 0.0003 AU \\
Orbital inclination, $i$ & 82.5 $\pm$ 0.5 {deg.} \\
Orbital eccentricity, $e$ & 0.0 (adopted) \\
Planet mass, $M_{\rm P}$ & 1.02 $\pm$ 0.03 $M_{\rm Jup}$ \\
Planet radius, $R_{\rm P}$  & 1.32 $\pm$ 0.08  $R_{\rm Jup}$ \\
$\log g_{\rm P}$ (cgs) & 3.12 $\pm$ 0.05 \\
Planet density, $\rho_{\rm P}$ & 0.44 $\pm$ 0.08 $\rho_{\rm Jup}$ \\
Planet temperature, $T_{\rm eql}$ & 1660 $\pm$ 40 K \\
\hline
\end{tabular} 
\end{table} 

\section{WASP-26 and the companion star}
\label{WASP26-double}

The visual double were investigated to determine whether they are physically
associated or just an optical double. Comparing the Palomar Observatory Sky
Survey (POSS-I) plates from the 1950s with more recent 2MASS images, shows that
there has been very little change in the separation and position angle of the
two stars over a period of some 50 years. This suggests that the system is at
least a common proper motion pair. The only proper motion measurements available
for the companion are those given in the UCAC3 catalogue
\citep{2010arXiv1003.2136Z}. This lists the proper motions of WASP-26 as
$\mu_{\rm RA} = +25.3 \pm 1.6$~mas\,y$^{-1}$ and $\mu_{\rm Dec} = -26.4 \pm
1.6$~mas\,y$^{-1}$, while those of the companion are given as $+114.6 \pm
9.1$~mas\,y$^{-1}$ and $-138.9 \pm 8.7$~mas\,y$^{-1}$, respectively. The
catalogue cautions that the values for the companion star are based on only two
position measurements and, thus, may not be reliable. In fact, if correct, the
UCAC3 proper motions would imply a change in separation of several arcseconds
over 50 years. This is clearly not supported by the survey images.

\begin{figure}
\includegraphics[height=\columnwidth,angle=-90]{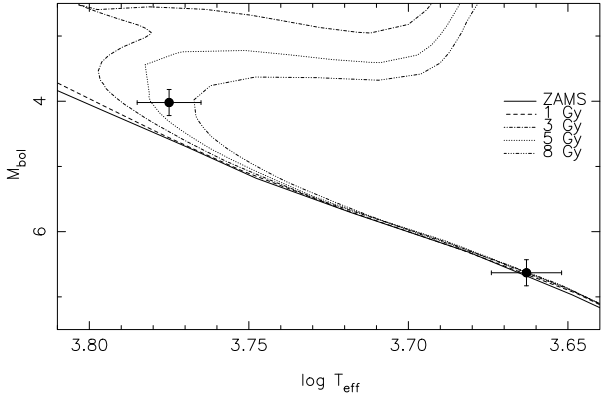}

\caption{$T_{\rm eff}$--$M_{\rm bol}$ diagram for WASP-26 and its companion
star. Various isochrones from \citet{2008A&A...482..883M} are given with ages
indicated in the figure.}

\label{HR-Diag}
\end{figure}

Using archival catalogue broad-band photometry from TYCHO, NOMAD, CMC14, DENIS
and 2MASS, bolometric magnitudes ($m_{\rm bol}$) for WASP-26 and the companion
star are estimated to be $11.0 \pm 0.1$ and $13.6 \pm 0.2$, respectively. Using
the Infrared Flux Method (IRFM) \citep{1977MNRAS.180..177B} and the 2MASS
magnitudes, the $T_{\rm eff}$ and angular diameter ($\theta$) of the two stars
are found to be: $T_{\rm eff} = 6010 \pm 140$~K and $\theta = 0.047 \pm
0.002$~mas, and $T_{\rm eff} = 4600 \pm 120$~K and $\theta = 0.024 \pm
0.001$~mas, respectively. The IRFM $T_{\rm eff}$ for WASP-26 is in good
agreement with that obtained from the spectroscopic analysis
(Table~\ref{WASP26-params}). A temperature--absolute bolometric magnitude
($T_{\rm eff}$--$M_{\rm bol}$) diagram for the two stars was constructed using
the evolutionary models of \citet{2008A&A...482..883M} (Fig.~\ref{HR-Diag}). A
distance modulus of $7.0 \pm 0.1$ ($\equiv 250 \pm 15$~pc) was required to bring
the companion star on to the main sequence. This is in good agreement with a
distance of $265 \pm 16$~pc to WASP-26 obtained using the radius determined from
MCMC analysis (Sect.~\ref{MCMC}). Hence, WASP-26 has evolved off the ZAMS with
an age of around $6 \pm 2$~Gy, which is in agreement with that estimated from
both lithium-depletion and gyrochronology (see Sect.~\ref{WASP-params}). 

There is clear evidence that the two stars appear to be physically associated. 
At a distance of 250~pc, the 15{\arcsec} projected-separation on the sky
corresponds to a physical separation of at least 3800~AU, implying an orbital
period of more than 170\,000~years.

\section{Conclusion}

WASP-26b is a moderately over-sized Jupiter-mass exoplanet transiting a G0 host
star every 2.7566~days. A simultaneous fit to transit photometry and
radial-velocity measurements gave a planetary mass of $1.02 \pm 0.03$~$M_{\rm
Jup}$ and radius of $1.32 \pm 0.08$~$R_{\rm Jup}$. The mass and radius of
WASP-26b place it within the group of bloated hot Jupiters. The incident flux
received by WASP-26b is $1.8 \times 10^9$~erg\,s$^{-1}$\,cm$^{-2}$, which is
clearly within the theoretical `pM' planetary class proposed by
\citet{2008ApJ...678.1419F}.

The host star, WASP-26, and its K-type companion are at least a common-proper
motion pair with a common distance of around $250 \pm 15$~pc and an age of
approximately $6 \pm 2$~Gy. With a physical separation of at least 3800~AU, the
companion star is unlikely to have any significant influence on the planetary
system's dynamics \citep{2007A&A...462..345D}. Spectroscopic confirmation of the
stellar parameters of the K-type companion should further improve the distance
and age determination of this system.

\section*{Acknowledgments}

The WASP Consortium comprises astronomers primarily from the Universities of
Keele, Leicester, The Open University, Queen's University Belfast, the
University of St Andrews, the Isaac Newton Group (La Palma), the Instituto de
Astrof\'isica de Canarias (Tenerife) and the South African Astronomical
Observatory (SAAO). WASP-South is hosted by SAAO and their support and
assistance is gratefully acknowledged. M. Gillon acknowledges support from the
Belgian Science Policy Office in the form of a Return Grant. This research has
made use of the VizieR catalogue access tool, CDS, Strasbourg, France. This
publication makes use of data products from the Two Micron All Sky Survey, which
is a joint project of the University of Massachusetts and the Infrared
Processing and Analysis Center/California Institute of Technology, funded by the
National Aeronautics and Space Administration and the National Science
Foundation.

\bibliographystyle{aa}

\end{document}